\documentclass[aps,pra,reprint,superscriptaddress,showpacs]{revtex4-1}
\usepackage{graphicx,amsmath,amssymb,ulem}
\usepackage{verbatim}
\usepackage{ragged2e}
\usepackage[dvipsnames]{xcolor}

\begin{document}
\title{Properties of bright squeezed vacuum at  increasing brightness}
\author{P.~R.~Sharapova}
\affiliation{Department of Physics, University of Paderborn, Warburger Stra\ss{}e 100, Paderborn D-33098, Germany}
\author{G.~Frascella}
\affiliation{Max-Planck Institute for the Science of Light, Staudtstr. 2, Erlangen  D-91058, Germany}
\affiliation{University of Erlangen-N\"urnberg, Staudtstr. 7/B2, 91058 Erlangen, Germany }
\author{M.~Riabinin}
\affiliation{Department of Physics, University of Paderborn, Warburger Stra\ss{}e 100, Paderborn D-33098, Germany}
\author{A.~M.~P\'erez}
\affiliation{Max-Planck Institute for the Science of Light, Staudtstr. 2, Erlangen  D-91058, Germany}
\affiliation{University of Erlangen-N\"urnberg, Staudtstr. 7/B2, 91058 Erlangen, Germany }\author{O.~V.~Tikhonova}
\affiliation{Physics Department, Moscow State University,  Leninskiye Gory 1-2, Moscow 119991, Russia}
\affiliation{Skobeltsyn Institute of Nuclear Physics, Lomonosov Moscow State University, Moscow 119234, Russia}
\author{S.~Lemieux}
\affiliation{Department of Physics,
University of Ottawa, 25 Templeton Street, Ottawa, Ontario K1N 6N5, Canada}
\author{R.~W.~Boyd}
\affiliation{Department of Physics,
University of Ottawa, 25 Templeton Street, Ottawa, Ontario K1N 6N5, Canada}
\affiliation{Institute of Optics, University of Rochester, Rochester, New York 14627, USA}
\author{G.~Leuchs}
\affiliation{Max-Planck Institute for the Science of Light, Staudtstr. 2, Erlangen  D-91058, Germany}
\author{M.~V.~Chekhova}
\affiliation{Max-Planck Institute for the Science of Light, Staudtstr. 2, Erlangen  D-91058, Germany}
\affiliation{University of Erlangen-N\"urnberg, Staudtstr. 7/B2, 91058 Erlangen, Germany }
\affiliation{Physics Department, Moscow State University, Leninskiye Gory 1-2, Moscow 119991, Russia}

\begin{abstract}
Bright squeezed vacuum (BSV) is a non-classical macroscopic state of light, which can be generated  through high-gain parametric down-conversion or four-wave mixing. Although BSV is an important tool in quantum optics and has a lot of applications, its theoretical description is still not complete. In particular, the existing description in terms of Schmidt modes fails to explain the spectral broadening observed in experiment as the mean number of photons increases. On the other hand, the semi-classical description accounting for the broadening does not allow to decouple the intermodal photon-number correlations. In this work, we present a new generalized theoretical approach  to describe the spatial properties of multimode BSV. In the multimode case one has to take into account the complicated interplay between all involved modes: each plane-wave mode interacts with all other modes, that complicates the problem significantly. The developed approach is based on  exchanging the $(\textbf{k},t)$ and $(\omega,z)$ representations and solving a system of integro-differential equations. Our approach predicts correctly the dynamics of the Schmidt modes and the broadening of the spectrum with the increase in the BSV mean photon number due to a stronger pumping. Moreover, the model  succesfully describes various properties of a widely used experimental configuration with two crystals and an air gap between them, namely an SU(1,1) interferometer. In particular, it predicts the narrowing of the intensity distribution, the reduction and shift of the side lobes, and the decline in the interference visibility as the mean photon number increases due to stronger pumping. The presented experimental results confirm the validity of the new approach. The model can be easily extended to the case of frequency spectrum, frequency Schmidt modes and other experimental configurations.

\end{abstract}
\maketitle

\section{Introduction}
At high  parametric gain, parametric down-conversion (PDC) and four-wave mixing (FWM) generate bright squeezed vacuum (BSV). BSV is a macroscopic non-classical state of light with  strong photon-number correlations (twin-beam squeezing)~\cite{Jedr, Bondani, Brida,Agafonov},  quadrature squeezing \cite{Corzo}, multimode structure \cite{Law, OptLett, Ou} and polarization entanglement if the generated photons have orthogonal polarizations \cite{polarizent}. BSV is  a promising tool for a lot of applications in quantum optics and metrology: imaging \cite{Lugiato, Boyer, Boyer1, BridaIm, AlleviIm},  quantum state engineering \cite{Our, Harder}, non-linear interferometry \cite{Yurke, review, exp}, super resolution and phase sensitivity  beyond the  shot-noise limit~\cite{Bridametr, Manceau}. Due to the high  mean photon number and the large number of modes involved, a theoretical description of BSV is complicated.

Several works on the theoretical description of multimode BSV are based on the coupled differential equations for the signal and idler plane-wave operators under the plane-wave pump approximation \cite{Kolobov, Klyshko, Dayan, Brambilla}. This approximation leads to the fact that each plane-wave signal mode interacts with only one plane-wave idler mode. This simplifies the problem significantly and allows to obtain analytical expressions for the output operators.  In other works, equations similar to the classical propagation equations were derived and their solutions based on the Green-function method were suggested \cite{Wasilewski}. Integro-differential equations for PDC with a fixed  spectral profile of the pump were written in \cite{Boyd, Christ,Eckstein}. The broadband-mode approach for the temporal domain based on the independent Schmidt modes was introduced in \cite{Christ} and developed in \cite{Eckstein}; the Schmidt-mode approach for the spatial domain was  developed and applied to experiment in \cite{SharapovaPRA}.   This approach describes several BSV effects and is very convenient for the analytical treatment of the problem. However, the Schmidt-mode theory neglects the energy mismatch between the pump, signal and idler photons and cannot describe the broadening of the spectrum, which is observed in experiment~\cite{Spasibko} as the BSV gets brighter due to the increase in the parametric gain (stronger pumping).

Moreover,  the behaviour of BSV properties with the increase of the parametric gain can be completely different depending on the geometry of experiment. For PDC in a single crystal, the spatial intensity distribution has a typical `sinc-squared' shape at low parametric gain, and it broadens  as the parametric gain increases (see the results below). Similar behavior was observed for the frequency spectrum in Ref.~\cite{Spasibko}. In contrast, a two-crystal configuration with an air gap in between leads to a complicated interference pattern of intensity with side lobes, which gets narrower with increasing parametric gain \cite{exp, frequency, SharapovaPRA}. This configuration is known as the SU(1,1) interferometer~\cite{Yurke} and it has recently attracted a lot of attention due to its  metrological applications~\cite{Hudelist,Gupta, Ma, Manceau,Shaked}.

In this work, we present a new theoretical approach to the description of the spatial properties of multimode BSV taking into account  the mode correlations. Our approach is based on exchanging the $(\textbf{k},t)$ and $(\omega,z)$ representations and solving the  high-dimensional system of integro-differential equations for plane-wave operators. This  approach describes various features of BSV, such as the intensity distribution and the shapes of the Schmidt modes, as well as their evolution with increasing parametric gain, both in the case of a single crystal and in the case of a two-crystal configurations. In full agreement with the experiment, the theory predicts the broadening of both the intensity distribution and the Schmidt-mode shapes with increasing gain in the case of a single crystal and the reduction of the side lobes in the two-crystal configuration. The suggested approach does not include any limitations on the pump waist width and the number of modes, as it was the case in the previous considerations, and it does not assume that the Hamiltonian commutes with itself at different moments of time.

The paper is organized as follows. Section~\ref{sec:single} describes the theoretical approach and applies it to the case of high-gain PDC in a single crystal. The Schmidt modes at variable parametric gain are considered in Section~\ref{sec:schmidt}. Section~\ref{sec:two} deals with the two-crystal configuration. In all sections, experimental results are also presented and compared with the theory. Finally, Section~\ref{sec:concl} is the conclusion.

\section{\label{sec:single} HIGH-GAIN PDC IN A SINGLE CRYSTAL}
The Hamiltonian of PDC in a crystal with a quadratic susceptibility $\chi^{(2)}(\mathbf{r})$  is given by \cite{Klyshko}
\begin{equation}
H\sim\int \mathrm{d}^3 r \chi^{(2)}(\mathbf{r})E_p^{(+)}(\mathbf{r},t)\hat{E}_s^{(-)}(\mathbf{r},t)\hat{E}_i^{(-)}(\mathbf{r},t)+\mathrm{h.c.},
\label{Ham}
\end{equation}
where $\hat{E}_{s,i}$ are electromagnetic field operators for signal/ idler photons, the pump is assumed to be a classical beam with a Gaussian envelope, propagating along the z axis, $E_p^{(+)} (\mathbf{r},t)= E_0 e^{-\frac{x^2+y^2}{2\sigma^2}}e^{i( k_p z-\omega_p t)}$,  with the full width at half maximum (FWHM) of the intensity distribution being $2\sqrt{\ln 2}\sigma$. By using the  quantization of the electromagnetic field, $\hat{E}_{s,i}^{(-)} (\mathbf{r},t)=\int \mathrm{d}{\mathbf{k}_{s,i}} C_{\mathbf{k}_{s,i}} e^{-i( \mathbf{k}_{s,i} \mathbf{r}-\omega_{s,i} t)} a^\dagger_{\mathbf{k}_{s,i}}$,where $a^\dagger_{\mathbf{k}_{s,i}}$ are the creation plane-wave operators, $C_{\mathbf{k}_{s,i}}$ are the coefficients of the decomposition, the Hamiltonian becomes
\begin{eqnarray}
H=\frac{i\hbar\Gamma}{2 \pi}\iint \mathrm{d} \mathbf{k}_s  \mathrm{d} \mathbf{k}_i \mathrm{d}^3\mathbf{r} e^{-\frac{x^2+y^2}{2\sigma^2}}e^{i k_p z}e^{i (-
\mathbf{k}_s-\mathbf{k}_i) \mathbf{r}}\times
\nonumber\\
e^{i (\omega_s+\omega_i-\omega_p)t}a^\dagger_{\mathbf{k}_s}a^\dagger_{\mathbf{k}_i} +\mathrm{h.c.}.
\label{Ham1}
\end{eqnarray}
Here we neglect the dependence of the coefficients $C_{\mathbf{k}_{s,i}}$ on $\mathbf{k}_{s,i}$ and suppose that {the interaction strength $\Gamma$, involving $\chi^{(2)}(\mathbf{r})$, the pump field amplitude, and other parameters} is a constant. For simplicity we consider a 2D model, using only one transverse coordinate $x$. After integration over $x$ and substituting $\mathrm{d} \mathbf{k}_{s,i}= \mathrm{d} q_{s,i} \mathrm{d}k_{sz,iz}$, where $q_{s,i}$ are the transverse wavevectors and $k_{sz,iz}$ are the longitudinal wavevectors of signal (idler) radiation, the Hamiltonian can be represented in the form
\begin{eqnarray}
H=\frac{i\hbar\Gamma}{2 \pi} \iint  \mathrm{d} q_s \mathrm{d}k_{sz}   \mathrm{d} q_i \mathrm{d}k_{iz} \mathrm{d} z e^{-\frac{(q_s+q_i)^2 \sigma^2}{2}} e^{i (k_p-k_{sz}-k_{iz})z}
\nonumber\\
\times e^{i (\omega_s+\omega_i-\omega_p)t}a^\dagger_{s}(q_s, k_{sz},t) a^\dagger_{i}(q_i, k_{iz},t) +\mathrm{h.c.}\ \ \ \ \ \ \ \ .
\label{Ham2}
\end{eqnarray}
 
This Hamiltonian is written in the momentum-time $(\mathbf{k},t)$ representation. In this picture, the Heisenberg equation of motion for the signal plane-wave operators takes the form
\begin{eqnarray}
\frac{\mathrm{d} a_s (q_s, k_{sz},t)}{\mathrm{d}t}= \frac{\Gamma}{2 \pi} \int \mathrm{d} q_i \mathrm{d}k_{iz} \mathrm{d} z e^{-\frac{(q_s+q_i)^2 \sigma^2}{2}}\times
\nonumber\\
 e^{i (k_p-k_{sz}-k_{iz})z}e^{i (\omega_s+\omega_i-\omega_p)t}a^{\dagger}_{i}(q_i, k_{iz},t), \ \ \ \ \ 
\label{Heis}
\end{eqnarray}
and similarly for the idler operators.
The operators $a^{\dagger}_{s,i}(q_{s,i}, k_{sz,iz},t)$ and $a_{s,i}(q_{s,i}, k_{sz,iz},t)$ defined before are the slowly varying parts of creation and annihilation operators. The fast varying components are connected with the slowly varying parts as
\begin{eqnarray}
\overline{a}^{\dagger}_{s,i}(q_{s,i}, k_{sz,iz},t)=e^{i \omega_{s,i} t}a^{\dagger}_{s,i}(q_{s,i}, k_{sz,iz},t),
\nonumber\\
\overline{a}^{\dagger}_{s,i}(q_{s,i}, \tilde{z},\tilde{\omega}_{s,i})= a^{\dagger}_{s,i}(q_{s,i}, \tilde{z},\tilde{\omega}_{s,i})e^{-i k_{sz,iz} (\tilde{\omega}_{s,i},q) \tilde{z}}.
\label{fast_varying}
\end{eqnarray} 
The Fourier transformation allows one to pass from $(q,k_z, t)$ to $(q,z, \omega)$ representation. For example, for the fast varying part of the idler creation operator the Fourier transformation is 
\begin{equation}
\overline{a}^{\dagger}_{i}(q_i, k_{iz},t)=\frac{1}{2 \pi}\int \overline{a}^{\dagger}_{i}(q_i, \tilde{z},\tilde{\omega}_i)e^{i\tilde{\omega}_i t} e^{i k_{iz} \tilde{z}} \mathrm{d} \tilde{\omega}_i \mathrm{d} \tilde{z}.
\label{Fourier}
\end{equation}

 After substituting (\ref{Fourier}) and (\ref{fast_varying}) in (\ref{Heis}) and integrating  its left and right parts over the interaction time, equation (\ref{Heis}) takes the form
\begin{eqnarray}
a_s (q_s, k_{sz},t=\tau)-a_s (q_s, k_{sz},t=0)=
\nonumber\\
\frac{\Gamma}{(2 \pi)^2} \int \mathrm{d}q_i \  \mathrm{d}k_{iz} \mathrm{d}z \ \mathrm{d}t  \ \mathrm{d} \tilde{\omega}_i \mathrm{d} \tilde{z} e^{-\frac{(q_s+q_i)^2 \sigma^2}{2}} e^{i (k_p-k_{sz}-k_{iz})z}\times
\nonumber\\
e^{-i (\omega_s+\tilde{\omega}_i-\omega_p)t}  e^{i k_{iz} \tilde{z}} a^{\dagger}_i(q_i, \tilde{z},\tilde{\omega}_i)e^{-i k_{iz}(\tilde{\omega}_i,q) \tilde{z}}, \ \ \ \ \ 
\label{Heis1}
\end{eqnarray}
where time $\tau$ corresponds to the end of the interaction. 
 The operators at the final $a_s (q_s, k_{sz},t=\tau)$ and initial $a_s (q_s, k_{sz},t=0)$ moments of time describe the boundary conditions. Using the boundary conditions, we have an equality between the operators in the end (beginning) of the interaction and operators corresponding to further (previous) free propagation in the linear medium. Outside the nonlinear medium,  $\mathbf{k}$ and $\omega$ are connected by the dispersion law $\omega=\omega (\mathbf{k})$ such that $a_{s,i} (q_{s,i}, k_{sz,iz},t)$ does not depend on $t$ and $a_{s,i} (q_{s,i}, z,\omega_{s,i})$ does not depend on $z$. This is the reason for which the boundary conditions in different representations are connected: $a_s (q_s, k_{sz},t=\tau)\sim a_s (q_s, L,\omega_s)$, $a_s (q_s, k_{sz},t=0) \sim a_s (q_s, L=0,\omega_s)$ \cite{Klyshko}, where $L$ is the length of the non-linear medium.

In the right-hand part of Eq. (\ref{Heis1}), integration over time and longitudinal momentum leads to the $\delta - $ functions,
\begin{eqnarray}
\frac{1}{2 \pi}\int \mathrm{d}t\  e^{-i (\omega_s+\tilde{\omega}_i-\omega_p)t}=\delta(\tilde{\omega}_i- \omega_p +\omega_s),
\nonumber\\
\frac{1}{2 \pi} \int \mathrm{d}k_{iz} e^{i k_{iz}(z-\tilde{z})}=\delta(z-\tilde{z}),
\label{delta}
\end{eqnarray}
the first of them defines the idler frequency through  the signal and pump frequencies.

The $\delta - $ functions  allow one to take integrals over $\mathrm{d} \tilde{\omega}$ and $\mathrm{d} \tilde{z}$ and simplify Eq.~(\ref{Heis1}):
\begin{eqnarray}
a_s (q_s, L,\omega_s)-a_s (q_s, 0,\omega_s)=
\Gamma \int \mathrm{d}q_i \int^L_0 \mathrm{d}z e^{-\frac{(q_s+q_i)^2 \sigma^2}{2}} \times 
\nonumber\\
 e^{i (k_p-k_{sz}-k_{iz} (\omega_p-\omega_s))z}
 a^{\dagger}_i(q_i, z,\omega_p-\omega_s). \ \ \ \ \ 
 \label{Heis2}
\end{eqnarray}
Taking a derivative of the left and the right parts of Eq. (\ref{Heis2}) with respect to the length of the nonlinear medium $L$ leads to the coupled integro-differential equations for the signal/idler annihilation/creation operators, 
\begin{eqnarray}
\frac{\mathrm{d} a_s (q_s, L,\omega_s)}{\mathrm{d}L}=
\Gamma \int \mathrm{d}q_i e^{-\frac{(q_s+q_i)^2 \sigma^2}{2}} \nonumber
\\
\times e^{i \Delta k L} a^{\dagger}_i(q_i, L,\omega_p-\omega_s), \ \ \ \ \
\label{Heis3}
\end{eqnarray}
where $\Delta k = k_p-k_{sz}(\omega_s)-k_{iz} (\omega_p-\omega_s) $.
A similar equation can be written for the idler frequency,
\begin{eqnarray}
\frac{\mathrm{d} a^{\dagger}_i(q_i, L,\omega_p-\omega_s)}{\mathrm{d}L}=
\Gamma \int \mathrm{d}q_s e^{-\frac{(q_s+q_i)^2 \sigma^2}{2}} \nonumber
\\
\times e^{-i \Delta k L} a_s (q_s, L,\omega_s). \ \ \ \ \
\label{Heis4}
\end{eqnarray}
 The solution to the system of coupled integro-differential equations (\ref{Heis3}, \ref{Heis4}) can be found in the form 
 \begin{eqnarray}
a_s(q_s, L, \omega_s) =
a_s(q_s) + \int \mathrm{d}q'_s \eta (q_s, q'_s, L,\Gamma) a_s(q'_s)+ \nonumber
\\  \int \mathrm{d}q'_i \beta(q_s, q'_i, L,\Gamma) a^{\dagger}_i(q'_i), \ \ \ \ \nonumber
\\
a^{\dagger}_i(q_i, L, \omega_i) =
a^{\dagger}_i(q_i) + \int \mathrm{d}q'_i \eta^{*} (q_i, q'_i, L,\Gamma) a^{\dagger}_i(q'_i)+\nonumber
\\  \int \mathrm{d}q'_s \beta^{*}(q_i, q'_s, L,\Gamma) a_s(q'_s),  \ \ \ \ 
\label{Heis5}
\end{eqnarray}
where $\eta(q_s, q'_s, L,\Gamma)$, $\beta(q_s, q'_i, L,\Gamma)$ are functions of the transverse wavevectors, crystal length and the interaction strength, and $a_s(q_s) = a_s(q_s, L=0, \omega_s)$ , $a^{\dagger}_i(q_i) = a^{\dagger}_i(q_i, L=0, \omega_i)$ are the initial plane-wave operators.

Finally, by solving the system of integro-differential equations Eq. (\ref{Heis3}-\ref{Heis4}) in the form of Eq. (\ref{Heis5}), various characteristics of BSV can be found. For example, the  mean photon-number distribution over transverse wavevectors  is
  \begin{eqnarray}
   N_s(q_s)=\langle a^{\dagger}_s (q_s, L,\omega_s) a_s (q_s, L,\omega_s) \rangle =\nonumber
   \\
    \int \mathrm{d}q_i' |\beta(q_s, q_i', L,\Gamma)|^2 .
    \label{nphotons}
     \end{eqnarray}
     In what follows, we assume small angles of emission  $\theta_{s,i}$, so that $q_s\approx k_s \theta_s$, and consider angular intensity distributions.

To find the connection between the theoretical parameter $\Gamma$ and the measured experimental gain, we have modelled the single plane-wave mode by calculating the total photon number in the collinear direction $(q_s=q_i=0)$ from Eq. (\ref{nphotons}) as a function of $\Gamma$ and fitted this dependence   by the well-known dependence for the single-mode regime, $y=B\sinh^2(A\Gamma)$, where $A$ and $B$ are the fitting parameters. Then the  parametric gain is defined as $G=A\Gamma$. A similar procedure was performed in the experiment. Using a pinhole, the dependence of the total intensity in the collinear direction on the parametric gain was measured and fitted by the function  $y=B_e\sinh^2(A_e\sqrt{P})$, where $A_e$ and $B_e$ are the fitting parameters, $P$ is the pump power. In this case, the experimental gain is defined as $G_e=A_e\sqrt{P}$, and the theoretical and experimental gains have to coincide.

\begin{figure}[htb]
\begin{center}
\includegraphics[width=0.35\textwidth]{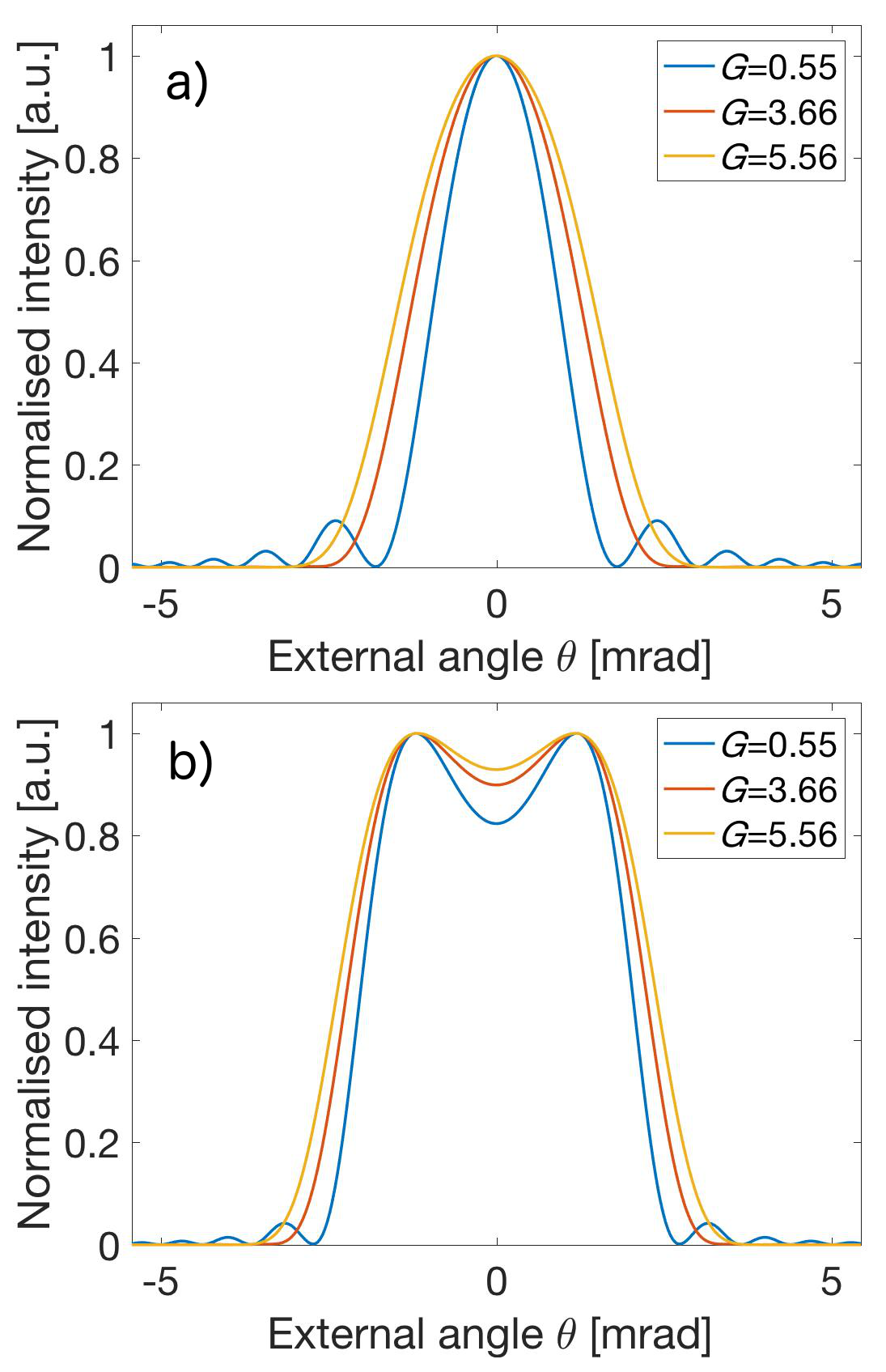}
\end{center}
\caption{The calculated normalized BSV intensity distributions for different  values of the parametric gain and phase mismatch $ \delta k  = 4530$ m$^{-1}$ (a) and $ \delta k  =- 3200$ m$^{-1}$ (b). } \label{fig:spectrum1}
\end{figure}

To compare the predictions of the described theory with the experiment, we considered a $2$ mm thick BBO crystal and a pump laser with the wavelength $354.7 $ nm and with a beam waist of FWHM $170$ $\mathrm{ \mu m}$. The transverse wavevector intensity distributions of type-I PDC were calculated using Eq. (\ref{nphotons}) for different  values of the parametric gain $G$, as shown in Fig. \ref{fig:spectrum1}. For ideal experimental conditions the phase mismatch $ \delta k = k_{p}(\theta_{axis}, \omega_{p}) - 2k_{s,i}(\omega_{s,i}) = 0$,  with $\theta_{axis}$ being the angle between the pump wavevector and the optic axis of the crystal.  Due to imperfect crystal alignment, the phase mismatch can be slightly non-zero. Such a deviation is hard to fix in the experiment but have an important effect on the shape of the intensity distribution, which is shown in panels (a) and (b) of Fig. \ref{fig:spectrum1}.

It is clearly seen that with increasing parametric gain the angular intensity distribution broadens. The broadening is directly connected with the fact that each signal plane-wave operator is coupled with all idler plane-wave operators in integro-differential equations through the functions $\beta$ and $\eta$ dependent on $\Gamma$. 

In experiment, BSV was obtained through PDC pumped by the third harmonic radiation (wavelength $354.7$ nm) of a pulsed Nd:YAG laser. The pulsed radiation (pulse width $18 $ ps, repetition rate $1$ kHz) is essential to reach the high-gain regime. The intensity distributions  were recorded with a charge-coupled device (CCD) camera in the Fourier plane of a lens with the focal length of $100 $ mm.
The spectral filtering  was performed using a band-pass filter with the transmission centered around the wavelength $710$ nm and with a bandwidth of $10 $ nm.

\begin{figure}[htb]
\begin{center}
\includegraphics[width=0.4\textwidth]{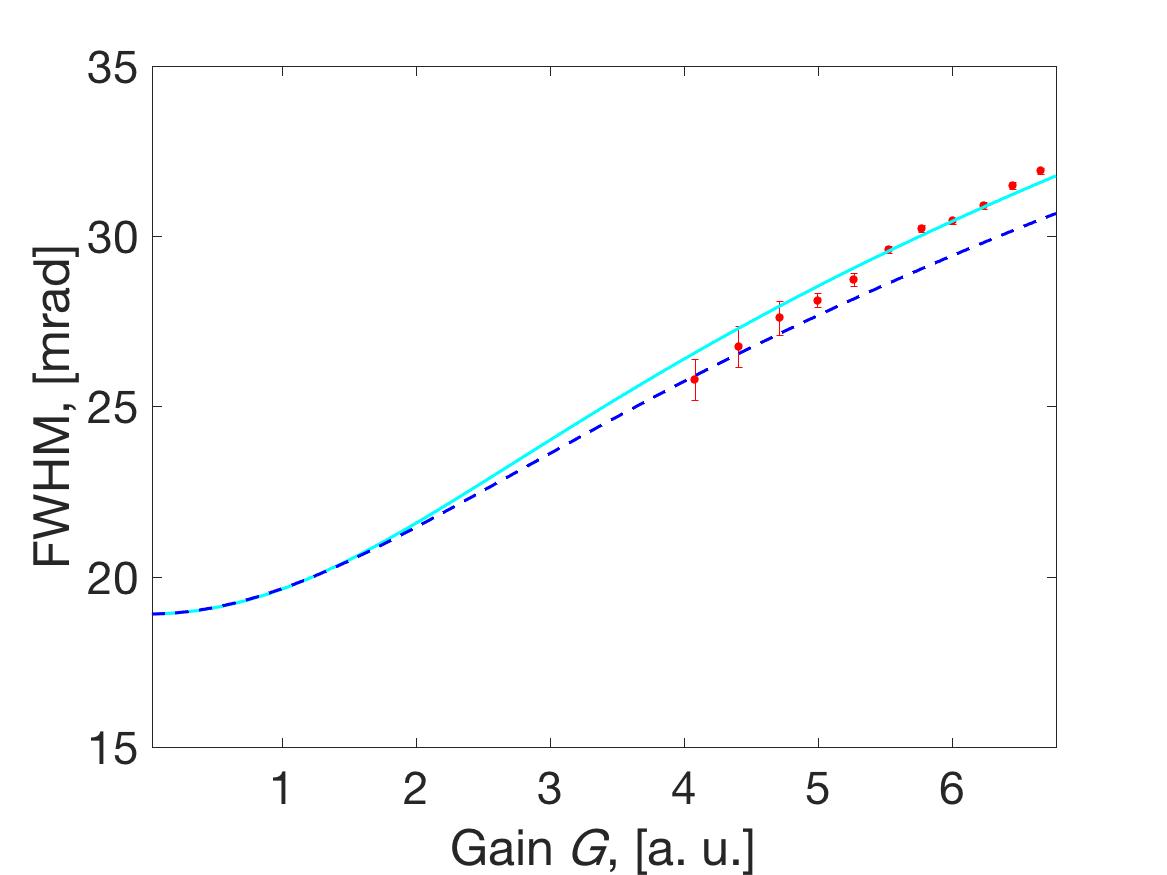}
\end{center}
\caption{The theoretical FWHM of the BSV intensity distribution vs parametric gain for a $170$ $\mathrm{\mu m}$ FWHM pump (cyan solid line) and plane-wave pump (blue dashed line) calculated according to the approach of Refs. \cite {Klyshko, Dayan}, and the experimental data (red dots).} \label{fig:FWHM}
\end{figure}

The dependence of the FWHM of the spatial intensity distribution of BSV on the parametric gain for $ \delta k  = 4530$ m$^{-1}$  (we fixed this value in further calculations according to the small mismatch in the experiment) was calculated and compared with the experimental data, see Fig. \ref{fig:FWHM}. In the low-gain regime, the FWHM of the BSV intensity distribution coincides with the value calculated using the first-order perturbation theory~\cite{Burlakov}. As the parametric gain increases, the FWHM monotonically grows. The same tendency is observed in the experiment (red points) and is in good agreement with the theoretical dependence (cyan line). The blue dashed line, calculated for the case of a plane-wave pump, predicts a slower broadening of the spectrum than the one with a focused pump.

\section{\label{sec:schmidt} SCHMIDT MODES}

The BSV radiation is strongly multimode. This multimode structure is important for a lot of applications [11] but, at the same time, is difficult to analyze. The most useful way to describe the multimode BSV radiation is by introducing a system of normalized orthogonal Schmidt modes. Within the Schmidt mode basis, each signal mode is only correlated with a single matching idler mode, which greatly facilitates the analysis.

In a simplified Schmidt-mode approach~\cite{SharapovaPRA}, the shapes of the BSV Schmidt modes do not depend on the parametric gain. In addition, the natural mode competition mechanism, i.e. low-order modes acquiring larger weights at larger gain, leads to the reduction in the number of Schmidt modes with increasing the gain. These two statements lead to the spatial (and frequency) narrowing of the intensity distribution, which is  in contradiction with the observed  broadening. The broadening can be only understood in the framework of the new approach considered here, and, as shown below,  is connected with the mode widths changing with the gain.



The complex function $\beta$ in Eq. (\ref{Heis5}) can be  written using the Schmidt decomposition with respect to the transverse wavevectors,

\begin{eqnarray}
\beta(q_s, q'_i, L,\Gamma) = \sum_{n}\sqrt{\Lambda_{n}} e^{i \phi_n} u_{n}\left(q_s, \Gamma \right) \psi_{n}\left(q'_i, \Gamma \right), \ \ \
\label{Schm_beta}
\end{eqnarray}
where gain-dependent eigenfunctions  $u_{n}(q_s, \Gamma )$ and $\psi_{n}(q'_i, \Gamma )$ for the signal and idler  beams, respectively,  are labeled with the index $n$, while $\Lambda_{n}$ represent gain-dependent weights of the decomposition, and $\phi_n$ are constant phases. Similarly, the function $\eta$ can be decomposed with eigenvalues $\tilde{\Lambda}_{n}$, constant phases $\varphi_n$, and eigenfunctions $v_{n}\left(q_s, \Gamma \right)$ and $\xi_{n}\left(q'_s, \Gamma \right)$:
\begin{eqnarray}
\eta (q_s, q'_s, L,\Gamma) = \sum_{n}\sqrt{\tilde{\Lambda}_{n}} e^{i \varphi_n} v_{n}\left(q_s, \Gamma \right) \xi_{n}\left(q'_s, \Gamma \right). \ \ \
\label{Schm_eta}
\end{eqnarray} 
From direct numerical decomposition of  $\beta(q_s, q'_i, L,\Gamma)$ and $\eta (q_s, q'_s, L,\Gamma)$ the following relations between eigenfunctions are ensued: $v_n(q, \Gamma)=u_n(q,\Gamma)$, $\xi^{*}_n(q,\Gamma)=\psi_n(q,\Gamma)$, and the absolute values of all functions are equal: $ |\psi_n|=|u_n|=|\xi_n|=|v_n|$. The gain-dependent weights $\Lambda_n$ and $\tilde{\Lambda}_n$ are different in the low-gain regime ($\Lambda_n > \tilde{\Lambda}_n$) but are getting closer with increasing the gain and become equal in the high-gain regime. Using the Schmidt decompositions in Eqs.  (\ref{Schm_beta}, \ref{Schm_eta}), we can introduce new photon creation/annihilation operators for the collective spatial Schmidt modes (the Schmidt-mode operators) of the  radiation as a result of the nonlinear interaction,
\begin{eqnarray}
   A^{\dagger}_n=\int \mathrm{d}q_s \xi^{*}_{n}\left(q_s,\Gamma\right)a^{\dagger}_s (q_s),\nonumber
   \\
   B^{\dagger}_n=\int \mathrm{d}q_i \psi_{n}\left(q_i,\Gamma\right)a^{\dagger}_i (q_i).
    \label{Schm_op}
     \end{eqnarray}
The operators $A^{\dagger}_n$ and $B^{\dagger}_n$ have the same form but they are related with the signal and idler plane-wave creation operators, respectively. Equations~(\ref{Heis5}) can be written in terms of the Schmidt operators, 
\begin{eqnarray}
a_s(q_s, L, \omega_s) =
a_s(q_s) + \sum_n u_n(q_s, \Gamma) (\sqrt{\tilde{\Lambda}_{n}}  e^{i \varphi_n} A_n\ \ \ \ \ \ \ \ \  \nonumber \\
+ \sqrt{\Lambda_{n}}e^{i \phi_n} B^{\dagger}_n),  \ \ \ \ \nonumber \\
a^{\dagger}_i(q_i, L, \omega_i) =
a^{\dagger}_i(q_i) + \sum_n v^{*}_n(q_i, \Gamma) (\sqrt{\tilde{\Lambda}_{n}} e^{-i \varphi_n} B^{\dagger}_n \ \ \ \ \ \ \ \ \   \nonumber  \\
+ \sqrt{\Lambda_{n}}e^{-i \phi_n}A_n).  \ \ \   \ \ \
\label{Schm}
\end{eqnarray}
Equations~(\ref{Schm}) clearly show that the output signal and idler plane-wave operators are connected with the same functions $u_n(q,\Gamma)=v_n(q,\Gamma)$, which is because we assumed frequency degeneracy.

The input/output relations~(\ref{Schm}) are similar to the ones of Ref.~\cite{SharapovaPRA}. However, in Eqs.~(\ref{Schm}) not only the weights $\Lambda_{n}$, $\tilde{\Lambda}_{n}$ but also the functions $u_n(q_s,\Gamma), v_n(q_i,\Gamma)$ depend on the parametric gain $\Gamma$. Thereby, the output operators are now defined by the Schmidt modes whose shapes are gain-dependent.
Using Eqs. (\ref{Schm}), the intensity distribution Eq. (\ref{nphotons}) can be written in a simple form as a sum of the squared absolute values of the Schmidt modes with  the corresponding weights:
 \begin{eqnarray}
   N_s(q_s)=\sum_{n}\Lambda_{n} |u_{n}\left(q_s,\Gamma\right)|^2.
     \end{eqnarray}

The Schmidt eigenmodes and eigenvalues of BSV can be reconstructed from the covariance of its intensity distribution~\cite{Beltran}. Indeed, consider the sum of the contributions of the signal and idler radiation for a fixed gain, i.e. $I_{\Sigma}\left(q\right)=I_{s}\left(q\right)+I_{i}\left(q\right)$, the covariance of intensities measured at positions $q$ and $q'$ is defined as
\begin{eqnarray}
\textrm{Cov}\left(q,q'\right)=\left<I_{\Sigma}\left(q\right)I_{\Sigma}\left(q'\right)\right>-\left<I_{\Sigma}\left(q\right)\right>\left<I_{\Sigma}\left(q'\right)\right>.
\label{covdef}
\end{eqnarray}
Calculation of the covariance distribution in terms of the Schmidt modes, using the input/output relations of Eqs. (\ref{Schm}), leads to
\begin{eqnarray}
\textrm{Cov}\left(q,q'\right)\propto
\left[\sum_{n}\Lambda_{n}u_{n}\left(q\right)u_{n}^{*}\left(q'\right)\right]^{2} \nonumber \\
+\left[\sum_{n}\Lambda_{n}v_{n}\left(q\right)v_{n}^{*}\left(q'\right)\right]^{2} \nonumber \\
+2\left|\sum_{n}\Lambda_{n}u_{n}\left(q\right)v_{n}\left(q'\right)\right|^{2},
\label{covschmidt}
\end{eqnarray}
where we suppose that $\Lambda_n = \tilde{\Lambda}_n$ for high gain.
The first two terms of Eq. (\ref{covschmidt}) are related to the auto-correlation of intensity fluctuations, respectively, of the signal and idler beams, while the third one represents the cross-correlation between the signal and idler radiation. 

For simplifying the reconstruction of the Schmidt modes, in experiment we eliminated the cross-correlations between the signal and idler radiation by filtering a wavelength slightly shifted from the degeneracy point, so that the detected signal photons did not have idler matches. In this case, in the covariance Eq. (\ref{covschmidt}) only the first term should remain, containing the signal Schmidt modes $u_n(q)$. Simultaneously, if the filtered wavelengths are still rather close to degeneracy, one can assume $u_n(q)=v_n(q)$. The Schmidt modes and weights were found by performing the singular value decomposition (SVD) of the square root of the covariance distribution~\cite{Finger}.
\begin{figure}[htb]
\begin{center}
\includegraphics[width=0.4\textwidth]{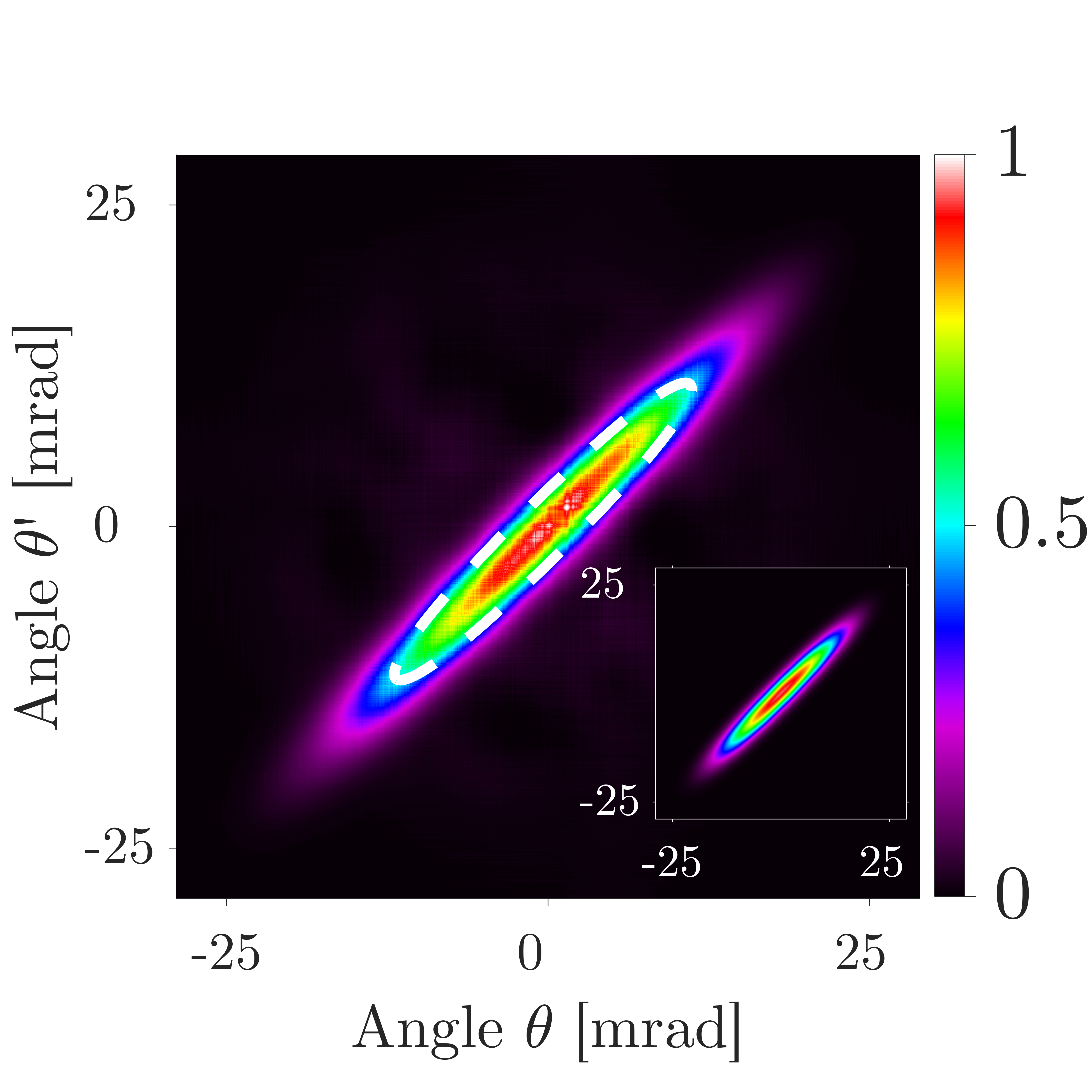}
\end{center}
\caption{Experimental distribution of the covariance in normalized units for the parametric gain $G_e=6$. Cross-correlations are removed by using a filter selecting only the signal radiation. For comparison, the white dashed line denotes the $0.5$ level of the fitted distribution, shown in the inset.} \label{fig:cov}
\end{figure}

In experiment, we filtered the signal radiation using a band-pass filter with the central wavelength $700 $ nm and a bandwith of $10 $ nm. Since the length of the nonlinear crystal was small enough ($2$ mm), we could neglect the effect of spatial walk-off. Accordingly,  the BSV radiation was axially symmetric and one could assume the factorability of the $x$ and $y$ degrees of freedom. Therefore, the analysis  was restricted to the intensity profiles along the $x$ direction. For a better signal to noise ratio we integrated the intensity distributions in the $y$ direction within the range from $- 2$ to $2$ mrad of the angle $q_y/\left|\mathbf{q}\right|$, where $\mathbf{q}=\left(q_x,q_y\right)$.  Around $2000$ single-shot intensity profiles $I\left(\theta \right)$ with $\theta=q_x/\left|\mathbf{q}\right|$  were measured, and the 2D covariance distribution $\textrm{Cov}\left(\theta,\theta'\right)$ was calculated. 

Figure~\ref{fig:cov} shows the experimental distribution of the covariance for $G_e=6$ in normalized units. The auto-correlation part is distributed along the main diagonal, where $\theta=\theta'$. Conversely, the cross-correlation, which would lead to  nonzero covariance values along the complementary diagonal $\theta=-\theta'$, is absent due to spectral filtering. In order to get rid of the background noise, a 2D fit was performed on the covariance distribution. Given the high-gain version of the intensity profile of the PDC radiation~\cite{Dayan} and the fact that the covariance distribution along the main diagonal behaves as the squared intensity profile, the function used for the fit was
\begin{eqnarray}
\textrm{Cov}\left(\theta,\theta' \right)=A+B \ e^{-C\left(\theta-\theta'\right)^2}\times \ \ \ \ \ \  \nonumber \\
\left[\frac{\sinh^{2}\sqrt{G_e^{2}-\left(D\left(\theta+\theta'\right)^{2}+E\right)^{2}}}{G_e^{2}-\left(D\left(\theta+\theta'\right)^{2}+E\right)^{2}}\right]^2,
\label{fitformula}
\end{eqnarray}
with $A, B, C, D$ being fitting parameters and $E$ being an experimentally determined quantity dependent on the phase mismatch. The inset of Fig.~\ref{fig:cov} demonstrates the fitted distribution for $G_e=6$, which is indeed equivalent to the experimental one. The white dashed line in the main figure represents the half-maximum level of the fitted covariance distribution and shows a good agreement with the half-maximum level (in cyan) for the experiment.

\begin{figure}[htb]
\begin{center}
\includegraphics[width=0.4\textwidth]{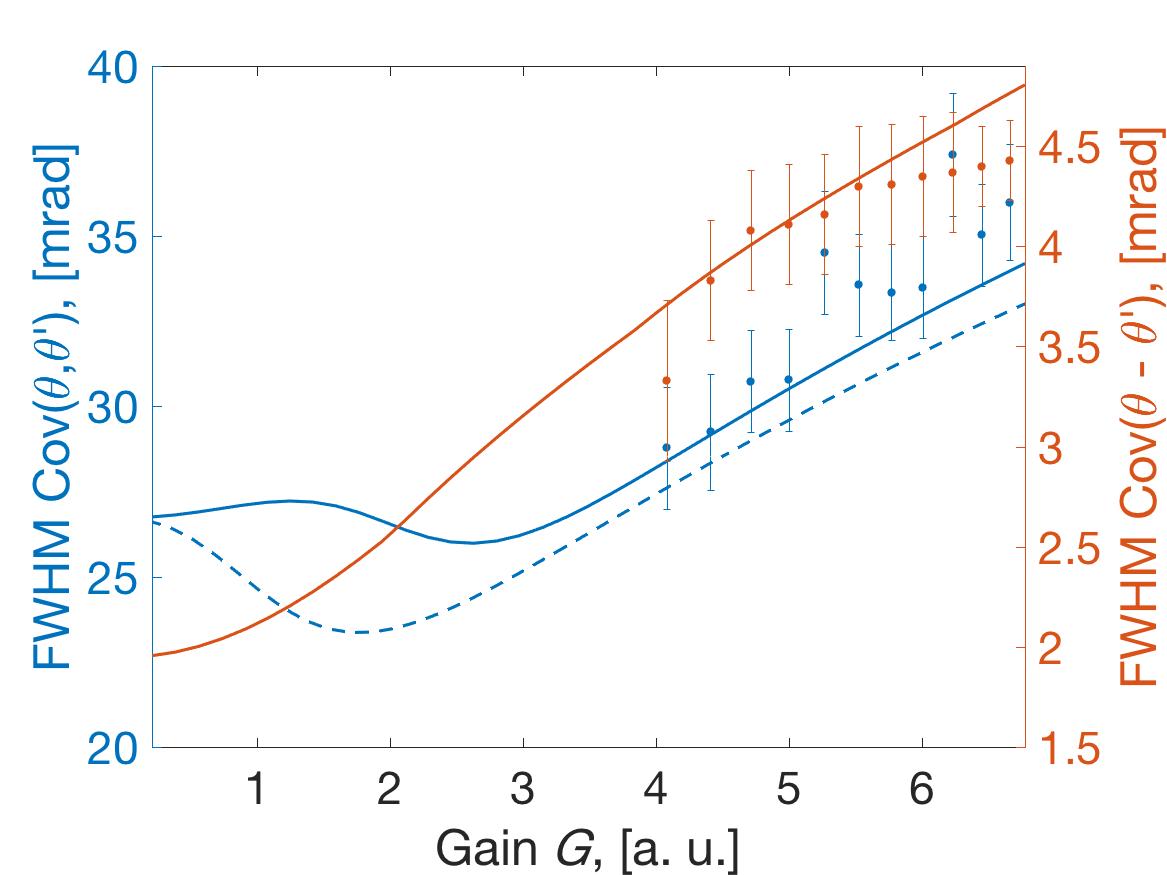}
\end{center}
\caption{Dependence of the FWHM of the covariance main (blue) and complementary (orange) diagonals on the parametric gain.  (Please note the different axis scales.) The solid lines represent theoretical calculations, while the points stand for the experimental data.
The dashed blue line corresponds to the main diagonal of covariance calculated under the plane-wave pump approximation \cite {Klyshko, Dayan}.}
\label{fig:FWHM covariance}
\end{figure}

\begin{figure}[htb]
\begin{center}
\includegraphics[width=0.45\textwidth]{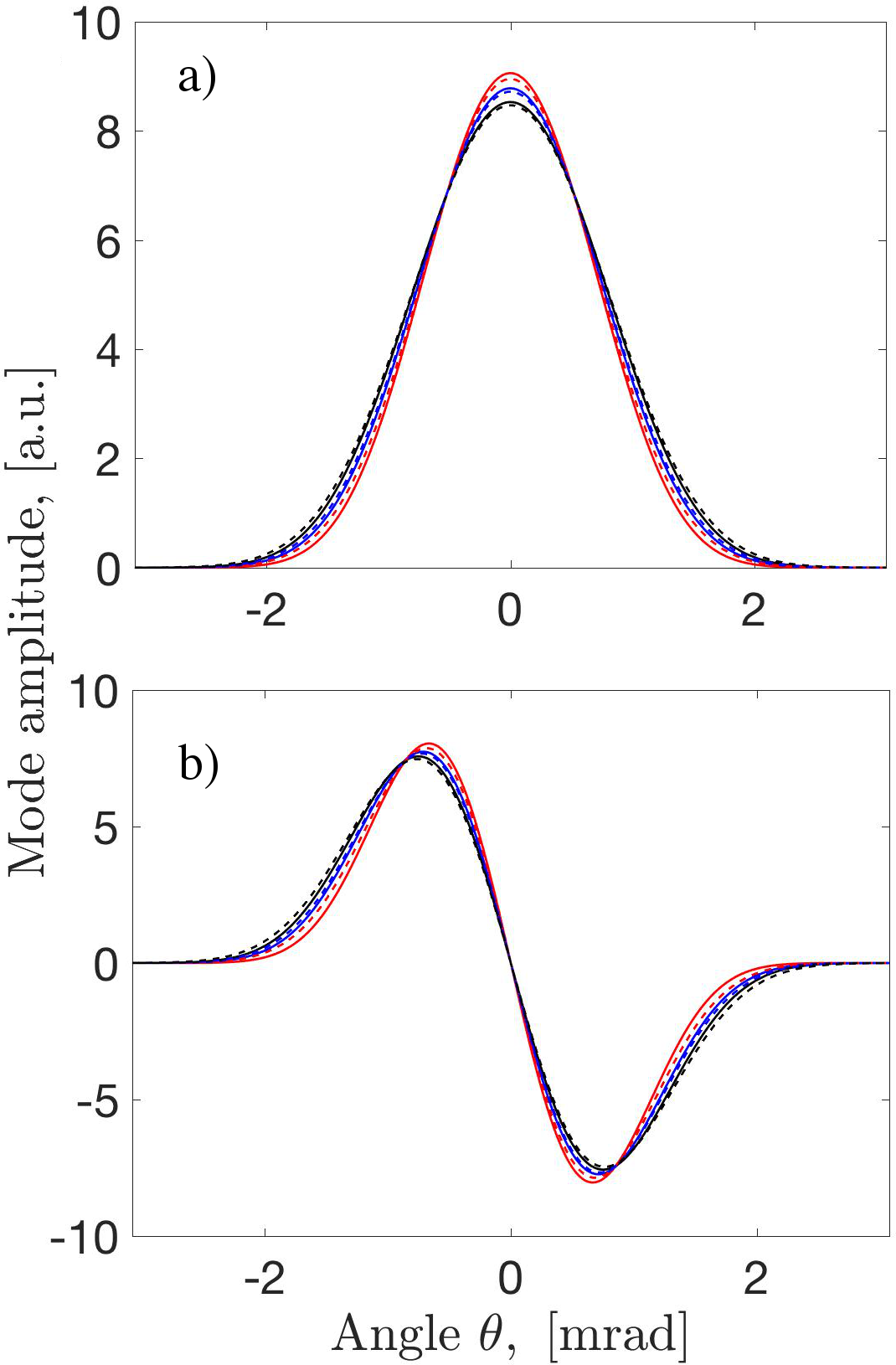}
\end{center}
\caption{a) The first and b) the second Shmidt modes for different parametric gain values: G=5.0 (red), G=5.7 (blue), G=6.6 (black). Solid lines represent theoretical calculations. Dashed lines stand for modes retrieved from the experiment. } \label{fig:modes together}
\end{figure}

The broadening of the covariance distribution with the increase of the parametric gain is shown in Fig. \ref{fig:FWHM covariance}. The experimental FWHM follows the predicted trend from the theory for both the main and the complementary diagonals. For low gain, the theoretical dependence coincides with the covariance obtained through the first-order perturbation theory. The dependence of the  main diagonal on the gain (blue solid curve in Fig. \ref{fig:FWHM covariance}) has a minimum. This minimum is also observed in the case of a plane-wave pump (blue dashed line in Fig. \ref{fig:FWHM covariance}) and qualitatively separates the low- and high-gain regimes.

The shapes and the weights of the Schmidt modes can be obtained through the SVD of the function $\sqrt{\textrm{Cov}\left(\theta,\theta'\right)}$. This procedure has been performed for both the theoretical, Eq.~(\ref{covschmidt}), and the fitted experimental covariance distributions. The results of the reconstruction for different gain values are shown in Fig.~\ref{fig:modes together} for the first and the second Schmidt modes. The shapes of the modes are close to the Hermite functions and their widths strictly depend on the gain. The general tendency is the broadening of the Schmidt modes with increasing gain, this broadening being more pronounced for  higher-order modes. The theoretical results show a good agreement with the experimental data.

\section{\label{sec:two} TWO-CRYSTAL CONFIGURATION}

The method described above can be extended to the two-crystal configuration, with the two crystals separated by an air gap of length $d$, known in the literature as the SU(1,1) interferometer~\cite{Yurke,OptLett,Hudelist,Gupta,Ma}. In this case, one should take into account that during the free propagation in the air gap the signal, idler and pump photons acquire an additional phase. This phase creates a factor of $ e^{i \Delta k' d}$ standing in front of the integral over the second crystal, where $\Delta k' = k^{air}_p-k^{air}_s-k^{air}_i$ is the wavevector mismatch in the air~\cite{Klyshkop}. Considering the problem step by step, we modify Eq.~(\ref{Heis1}) by taking into account the $\delta$ - function conditions in Eq. (\ref{delta}):

\begin{eqnarray}
a_s (q_s, k_{sz},t=\tau)-a_s (q_s, k_{sz},t=0)=
\nonumber\\
\Gamma \int \mathrm{d}q_i  e^{-\frac{(q_s+q_i)^2 \sigma^2}{2}}  
[\int^L_0 \mathrm{d}z e^{i \Delta k z}
 a^{\dagger}_i(q_i, z,\omega_p-\omega_s)+
 \nonumber\\
 e^{i \Delta k' d}\int^{2L}_L \mathrm{d}z e^{i \Delta k z} a^{\dagger}_i(q_i, z,\omega_p-\omega_s) ]. \ \ \
\label{2crystals}
\end{eqnarray}
The boundary conditions connect the beginning and the end of the interaction in different representations:  $a_s (q_s, k_{sz},t=0) \sim a_s (q_s, L=0,\omega_s)$,  $a_s (q_s, k_{sz},t=\tau) \sim a_s (q_s, 2L,\omega_s)$.

The derivative of Eq. (\ref{2crystals}) over the parameter $L$ gives 

\begin{eqnarray}
\frac{\mathrm{d} a_s (q_s, 2L,\omega_s)}{\mathrm{d}L}=
\Gamma \int \mathrm{d}q_i  e^{-\frac{(q_s+q_i)^2 \sigma^2}{2}} \times 
 \nonumber\\
 (e^{i \Delta k L} \ (1-e^{i \Delta k' d}) \
 a^{\dagger}_i(q_i, L,\omega_p-\omega_s)+
  \nonumber\\
+ 2 e^{i \Delta k' d} e^{2i \Delta k L} a^{\dagger}_i(q_i, 2L,\omega_p-\omega_s)),\ \ \ \ 
\label{2crystals_1}
\end{eqnarray}
and for the idler creation operator,
\begin{eqnarray}
\frac{\mathrm{d} a^{\dagger}_i (q_i, 2L,\omega_p-\omega_s)}{\mathrm{d}L}=
\Gamma \int \mathrm{d}q_s  e^{-\frac{(q_s+q_i)^2 \sigma^2}{2}} \times 
 \nonumber\\
 (e^{-i \Delta k L} \ (1-e^{-i \Delta k' d}) \
 a_s(q_s, L,\omega_s)+ 
 \nonumber\\
+ 2 e^{-i \Delta k' d} e^{-2i \Delta k L} a_s(q_s, 2L,\omega_s)),\ \ \ \ 
\label{2crystals_2}
\end{eqnarray}
where $a^{\dagger}_i(q_i, L,\omega_p-\omega_s)$ and $a_s(q_s, L,\omega_s)$ can be found by solving Eqs.~(\ref{Heis3}, \ref{Heis4}) for the single crystal.

\begin{figure*}[htb]
\begin{center}
\includegraphics[width=0.92 \textwidth]{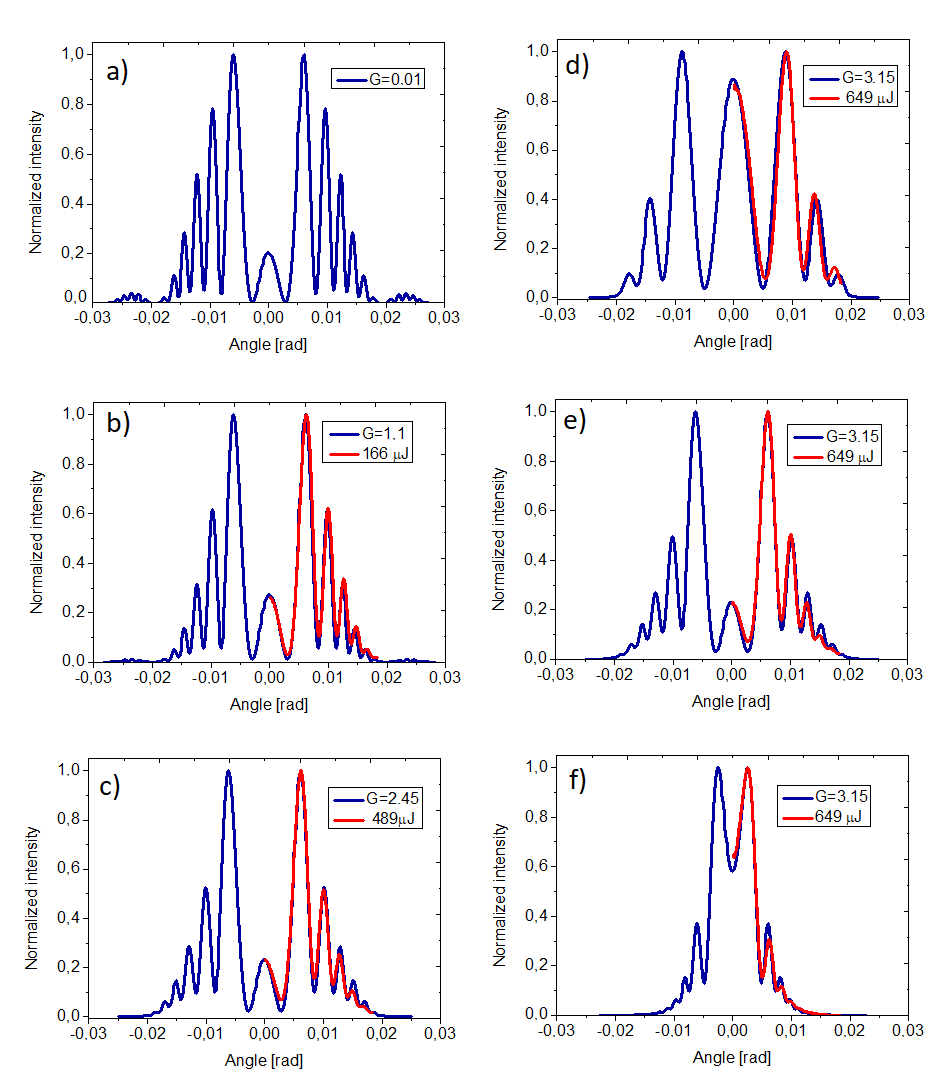}
\end{center}
\caption{ The normalized intensity distributions of BSV in the two-crystals configuration with an air gap. a)-c) The distance between the crystals is $10.66$ $\mathrm{mm}$, the gain in each crystal is a) $G=0.01$, b) $G=1.1$, c) $G=2.45$. d-f) The gain is fixed, $G=3.15$, the distance between the crystals is d) $5.58$ $\mathrm{mm}$,  e) $10.66$ $\mathrm{mm}$, f) $23.36$ $\mathrm{mm}$. Blue curves  are theoretical calculations, red curves present the experimental data. The legends also show the values of the pump energy per pulse.}
 \label{fig:distance}
\end{figure*}

The intensity distribution in the presence of the air gap is completely different from the single-crystal case and depends on the length of the air gap. Due to the different refractive indices of the pump and BSV photons in the air, an additional phase is acquired in the gap and the intensity of light emitted in the collinear direction oscillates from minimum to maximum as $d$ increases.   Also, an increase in $d$ leads to more and more frequent interference fringes in the intensity distribution (Fig.~\ref{fig:distance} a). 

In the experiment, we  pumped two nonlinear crystals with the third harmonic of a pulsed Nd:YAG laser (repetition rate 50\,Hz, wavelength 354.7\,nm, and pulse duration 29.4\,ps) with a FWHM diameter of approximately 0.3\,mm. 
The two BBO crystals (3-mm thick, cut for degenerate type-I PDC) were aligned, in turn, for degenerate phase matching. 
A dichroic mirror and a color-glass filter suppressed the pump after the nonlinear interaction. 
A band-pass filter selected a 10\,nm bandwidth (FWHM) of the PDC around the wavelength 710\,nm. 
A lens  brought the PDC to the momentum space, where a CMOS camera  was introduced. 
On the camera, the background was subtracted and the data was acquired for 200\,ms.
The pump energy per pulse was measured before the crystals with a calibrated energy meter. 
The distance between the crystals was varied by changing the position of the first crystal using a translation stage.
Neutral density filters  were used to avoid the saturation of the camera.
The ring patterns measured with the camera were then transformed into polar coordinates. 
The radial profiles were obtained by averaging out the polar plots over the azimuthal angle.

As the pump power increases, the parametric gain in each crystal grows (the gain after two crystals with a gap grows non-monotonically). Figure~\ref{fig:distance} a-c shows the resulting spectra both calculated (blue) and measured (red), with the parametric gain and the pump pulse energy shown in each panel. (In the low-gain case, the measurement was not possible because of the small intensity.)  Apart of a small shift in the fringes,
increasing the pump power leads to the reduction of the side peaks. Therefore, the envelope of the spectrum gets narrower as the  parametric gain grows, in contrast to the single-crystal case.  Moreover, from Fig. \ref{fig:distance} a-c it is clearly seen  that the visibility of the interference fringes drops down with increasing  pump power.  Note that with stronger pumping the Kerr effect, which  leads to an additional phase (mostly manifested in the collinear direction) \cite{Beltran}, becomes more pronounced and must be taken into account for the correct description.

In Fig. \ref{fig:distance} d-f, one can observe that with increasing the distance between the crystals, the total angular width of the spectrum for the same pump power (or for the same gain in each crystal) is reduced, as reported in \cite{SharapovaPRA}. This happens due to diffraction, which leads to the reduction of the angular width of BSV that overlaps with the pump and is amplified in the second crystal; diffraction is more pronounced for larger distances. As it was mentioned above, the second mechanism leading to the narrowing of the spatial intensity distribution in the two-crystal configuration is the increase of the pump power (the parametric gain in each crystal). Simultaneously, these two mechanisms diminish the number of the Schmidt modes in the system, allowing one to create different shapes of intensity distributions and to control the number of modes in BSV.

\section{\label{sec:concl} CONCLUSION}
We have presented a new theoretical approach to describe the spatial properties of BSV generated through high-gain PDC. In this approach, we derived and solved the  integro-differential equations for plane-wave operators without limitations on the pump waist width, number of modes and commutation of the Hamiltonian at different moments of time. The developed approach successfully captures a lot of features of BSV. On the one hand, it is compatible with the Schmidt-mode representation. On the other hand, it properly describes the broadening of the spectrum with increasing parametric gain. As a result, the new treatment correctly predicts the dependence of the Schmidt mode widths on the parametric gain. The model describes different experimental configurations: the single-crystal case and the configuration of two crystals with an air gap between them. For the verification of our theoretical model, we have performed several experiments, both with a single-crystal and with a two-crystals PDC source (SU(1,1) interferometer). The presented experimental results are in good agreement with performed theoretical calculations. Our model gives a deep insight into the properties of high-gain PDC, its mode structure, the origin of nonclassical correlations; opens a broad opportunities for profound investigation and in-depth understanding of BSV, development and implementation of its numerous unique applications.

\section{ACKNOWLEDGMENTS}
Financial support of the Deutsche Forschungsgemeinschaft(DFG) through TRR 142, Project No. C02 is gratefully acknowledged.
P.R.Sh. thanks the state of Nordrhein-Westfalen for support by the \textit{Landesprogramm f\"ur geschlechtergerechte
Hochschulen}.   We also acknowledge financial support of  the DFG project SH 1228/3-1 and joint DFG-Russian Science Foundation (RSF) project SH 1228/2-1, ME 1916/7-1 - No.19-42-04105.


\begin{thebibliography}{99}

\bibitem{Jedr} O. Jedrkiewicz, Y.-K. Jiang, E. Brambilla, A. Gatti, M. Bache, L. A. Lugiato, and P. Di Trapani, Phys. Rev. Lett. \textbf{93}, 243601 (2004).

\bibitem{Bondani} M. Bondani, A. Allevi, G. Zambra, M. G. A. Paris, and A. Andreoni, Phys. Rev. A \textbf{76}, 013833 (2007).

\bibitem{Brida} G. Brida, L. Caspani, A. Gatti, M. Genovese, A. Meda, and I. R. Berchera, Phys. Rev. Lett. \textbf{102}, 213602 (2009).

\bibitem{Agafonov}I. N. Agafonov, M. V. Chekhova, and G. Leuchs, Phys. Rev. A \textbf{82}, 011801(R) (2010).

\bibitem{Corzo} N. Corzo, A. M. Marino, K. M. Jones, and P. D. Lett, Opt. Exp. \textbf{19}, 21358 (2011).

\bibitem{Law} C. K. Law, I. A. Walmsley and J. H. Eberly Phys. Rev. Lett. \textbf{84}, 5304 (2000).

\bibitem{OptLett} A. M. P\'erez, T. Sh. Iskhakov, P. Sharapova, S. Lemieux, O. V. Tikhonova,M. V. Chekhova, and G. Leuchs, Opt. Lett. \textbf{39}, 2403 (2014).

\bibitem{Ou}X. Guo, N. Liu, X. Li, Z. Y. Ou, Opt. Exp. \textbf{23}, 29369 (2016).

\bibitem{polarizent}T. Sh. Iskhakov, I. N. Agafonov, M.V. Chekhova, and G. Leuchs, Phys. Rev. Lett. \textbf{109}, 150502 (2012).

\bibitem{Lugiato} L. A. Lugiato, A. Gatti and E. Brambilla, J. Opt. B: Quantum Semiclass. Opt. \textbf{4},  S176–S183, (2002).

\bibitem{Boyer} V. Boyer, A. M. Marino, R. C. Pooser, P. D. Lett, Science \textbf{321}, 544 (2008).

\bibitem{Boyer1}V. Boyer, A.M. Marino, and P. D. Lett, Phys. Rev. Lett. \textbf{100}, 143601 (2008).

\bibitem {BridaIm}G. Brida, M. Genovese, and I. R. Berchera, Nature Photonics \textbf{4}, 227 (2010). 

\bibitem {AlleviIm} A. Allevi and M. Bondani, J. Opt. \textbf{19}, 064001 (2017). 
 
\bibitem{Our} A. Ourjoumtsev, H. Jeong, R. Tualle-Brouri, and P. Grangier, Nature (London) \textbf{448}, 784 (2007). 

\bibitem{Harder} G. Harder, T. J. Bartley, A. E. Lita, S. W. Nam, T. Gerrits, and C. Silberhorn, Phys. Rev. Lett. \textbf{116}, 143601 (2016).
 
\bibitem{exp}S. Lemieux, M. Manceau, P. R. Sharapova, O. V. Tikhonova, R. W. Boyd, G. Leuchs and M. V. Chekhova, Phys. Rev. Lett. \textbf{117}, 183601 (2016).

\bibitem{Yurke}B. Yurke, S.L. McCall, and J.R. Klauder, Phys. Rev. A \textbf{33}, 4033 (1986).

\bibitem{review} M. V. Chekhova and Z. Y. Ou, Advances in Optics and Photonics \textbf{8}, 104 (2016).

\bibitem {Bridametr} G. Brida, I. P. Degiovanni, M. Genovese, M. L. Rastello, and I. R.-Berchera, Opt. Exp. 18, 20572 (2010). 

\bibitem {Manceau} M. Manceau, G. Leuchs, F. Ya. Khalili, and M.V. Chekhova, Phys. Rev. Lett. 119, 223604 (2017). 

\bibitem{Kolobov} M. I. Kolobov, Rev. Mod. Phys. \textbf{71}, 1539 (1999).

\bibitem{Klyshko} D. N. Klyshko, Photons and Nonlinear Optics. Gordon and Breach Science Publishers (1988).


\bibitem{Dayan} B. Dayan, Phys. Rev. A \textbf{76}, 043813 (2007).

\bibitem{Brambilla} E. Brambilla, L. Caspani, O. Jedrkiewicz, L. A. Lugiato, and A. Gatti, Phys. Rev. A \textbf{77}, 053807 (2008).

\bibitem{Wasilewski} W.~Wasilewski, A.~I.~Lvovsky, K.~Banaszek, and C.~Radzewicz, Phys. Rev. A \textbf{73}, 063819 (2006).

\bibitem{Christ} A. Christ, B. Brecht, W. Mauerer, and C. Silberhorn, New Journ. of Phys. \textbf{15}, 053038 (2013).

\bibitem{Eckstein} A. Eckstein, B. Brecht, and C. Silberhorn, Opt. Express \textbf{19}, 13770 (2011).

\bibitem{Boyd}R.~S.~Bennink and R.~W.~Boyd, Phys. Rev. A \textbf{66}, 053815 (2002).

\bibitem{SharapovaPRA} P. Sharapova, A. M. P\'erez, O. V. Tikhonova, and M. V. Chekhova, Phys. Rev. A \textbf{91}, 043816 (2015).

\bibitem{Spasibko} K. Yu. Spasibko, T. Sh. Iskhakov, and M. V. Chekhova, Spectral properties of high-gain parametric down-conversion. Opt. Exp. \textbf{20}, 7507 (2012). 

\bibitem{frequency}P. R. Sharapova, O. V. Tikhonova, S. Lemieux, R. W. Boyd, and M. V. Chekhova, Phys. Rev. A \textbf{97}, 053827 (2018).

\bibitem{Hudelist}F. Hudelist, J. Kong, C. Liu, J. Jing, Z. Y. Ou, and W. Zhang, Nature Comm. \textbf{5}, 3049 (2014).

\bibitem{Gupta}P. Gupta, B. L. Schmittberger, B. E. Anderson, K. M. Jones, and P. D. Lett, Opt. Exp. \textbf{26}, 391-401 (2018).

\bibitem{Ma}X. Ma, C. You, S. Adhikari, E. S. Matekole, R. T. Glasser, H. Lee, J. P. Dowling, Opt. Exp. \textbf{26}, 18492-18504 (2018).

\bibitem{Shaked}Y. Shaked, Y. Michael, R. Z. Vered, L. Bello, M. Rosenbluh, and A. Pe’er, Nature Comm. s41467-018-03083-5 (2018).

\bibitem{Burlakov}  A.~V.~Burlakov, M.~V.~Chekhova, D.~N.~Klyshko, S.~P.~Kulik, A.~N.~Penin, Y.~H.~Shih, and D.~V.~Strekalov, Phys. Rev. A \textbf{56}, 3214 (1997).

\bibitem{Klyshkop} D. N. Klyshko, Zh. Eksp. Teor. Fiz. \textbf{104}, 2676-268 (1993).

\bibitem{Beltran} L. Beltran, G. Frascella, A.~M. P\'erez, R. Fickler, P.~R.~Sharapova, M. Manceau, O.~V. Tikhonova, R.~W. Boyd, G. Leuchs, and  M.~V. Chekhova, J. Opt. \textbf{19}, no.~4,
  044005 (2017).

\bibitem{Finger} M. A. Finger, N. Y. Joly, P. St. J. Russell, and M. V. Chekhova, Phys. Rev. A \textbf{95}, 053814 (2017).





\end{thebibliography}
\end{document}